\documentstyle[11pt,paspconf,psfig]{article}

\begin{document}
\title{Continuum Observations of the High-Redshift Universe at 
          Sub-millimetre Wavelengths}

\author{David H. Hughes \altaffilmark{1} 
\& James S. Dunlop \altaffilmark{2}}
\affil{Institute for Astronomy, Department of Physics \& Astronomy, 
       University of Edinburgh, Royal Observatory, 
       Edinburgh EH9 3HJ, U.K.}

\altaffiltext{1}{d.hughes@roe.ac.uk}
\altaffiltext{2}{j.dunlop@roe.ac.uk}

\begin{abstract}
New bolometer arrays operating on the world's largest sub-millimetre and
millimetre telescopes offer a unique view of the high-redshift universe 
with unprecedented sensitivity. Recent sub-millimetre continuum 
studies show that the host galaxies of many luminous high-redshift 
active galactic nuclei (radio galaxies and radio-quiet quasars) 
radiate strongly at rest-frame far-infrared wavelengths and thus contain
substantial quantities of dust. In the majority of these high-redshift 
AGN-hosts, the inferred star formation is proceeding at a rate comparable 
to that found in local, interacting ultra-luminous far-infrared galaxies. 
This level of activity is an order of magnitude greater 
than the more modest star-formation rates apparently displayed by 
the recently-discovered Lyman-limit galaxies at ${\rm z} \sim 3$, which 
have been argued to represent the era of spheroid formation (although the 
degree to which the effects of reddening by dust grains may 
have biased the interpretation of these optical/UV studies of high-z
galaxies has yet to be properly determined). However, it is too early to
say whether such bright far-infrared emission is a feature of all massive
galaxies at $z > 3$, or whether it is in fact confined to the hosts of
the most luminous AGN.

In this paper we review the current status of cosmological
observations at sub-millimetre and millimetre wavelengths, highlighting
our own recent SCUBA observations of high-redshift radio galaxies.
We also explain how observations over the next few years should allow the 
true level of star-formation activity in the high-redshift universe to be 
properly quantified, and we provide example predictions 
for the first deep sub-millimetre survey (of the Hubble Deep Field), 
which we and our colleagues are currently undertaking 
with SCUBA at the JCMT.

\end{abstract}


\section{Introduction}

Sub-millimetre (sub-mm) cosmology is still in its infancy. Despite significant
efforts
with single-element bolometers over the last 5 years, only a handful of
unlensed objects have been unambiguously detected (- for a summary, see 
Hughes, Dunlop \& Rawlings 1997). 
However, the recent advent of sensitive sub-mm/mm
bolometer arrays seems set to revolutionize the field.
Studies of statistically significant samples of known high-redshift 
sources are now feasible, as are the first meaningful 
sub-mm blank-field surveys. 
The key logical steps in our own approach to this burgeoning field can be summarized as follows:

\begin{itemize}
\item Make pointed observations of known high-redshift
objects spanning a wide range in redshift. We have concentrated on 
high-redshift steep-spectrum radio galaxies primarily 
because they can be reasonably expected to be the progenitors of at 
least a subset of present-day massive ellipticals, but also because, being
selected on the basis of extended emission, their sub-mm 
properties are unlikely to be biased by gravitational lensing.

\item Use sub-mm/far-infrared(FIR) data to first identify the 
emission mechanism 
that dominates the production of the rest-frame sub-mm/FIR luminosity. 
Then, if this emission is proven to be optically-thin thermal re-radiation 
from dust grains, use these same data to 
constrain the dust temperature, and hence calculate a reddening-free measure 
of the dust mass and infer the total mass of molecular gas in the galaxy.  
Assuming the dust grains are heated by young, massive stars, and not by an AGN,
then estimate the `current' star-formation rate (SFR).

\item Compare the gas masses available for further star formation in 
the host galaxies of high-redshift active galactic nuclei (AGN), 
with the the final stellar masses of their expected low-redshift counterparts.
Hence infer the evolutionary status of the galaxies that host
high-redshift AGN.

\item Building on the results of such pointed observations, design and
undertake a series of complementary sub-mm 
surveys reaching different depths 
over different areas. Use the measured source counts and redshift distributions
to determine the true level and history 
of star-formation activity at high redshift,
and to determine the form of the cosmological evolution of the dust-enshrouded
starburst population.

\end{itemize}

\section{Bright far-infrared emission from massive galaxies at high-redshift}
In the local universe, powerful active galactic nuclei are found to reside 
exclusively in extremely luminous host galaxies. Near-infrared images
of the dominant stellar populations in the hosts of luminous 
low-redshift quasars and radio galaxies show that they have $K$-band
luminosities $\equiv 2 - 5 L^{\star}_{K}$ and that, morphologically, 
radio galaxies, the hosts of radio-loud quasars, and even the hosts of 
the more luminous radio-quiet quasars appear to be giant ellipticals 
(Taylor {\it et al.} 1996).
It is therefore natural to assume that the hosts of luminous high-redshift 
radio sources  are among the progenitors of the most massive 
present day ellipticals which have stellar masses
${\rm M_{stars} > 5 \times 10^{11}M_{\sun}}$.
Models for the photometric evolution of elliptical 
galaxies ({\it e.g.} Mazzei \& de Zotti 1996)
suggest that within the first few Gyr of their lifetimes 
a significant fraction ($\sim 30\%$) 
of their bolometric luminosity is radiated at rest-frame FIR 
wavelengths due to rapid massive star-formation proceeding at a rate
$> 100 {\rm M}_{\sun}/{\rm yr}$. 
At high redshifts, $z > 2$, this characteristic FIR ($60-200$\micron) 
spectral peak is shifted into the longer-wavelength sub-mm regime. It is 
the existence of this relatively brief luminous, starburst phase in 
the early evolution of massive galaxies that we aim to either confirm 
or refute.

Sub-mm observations may well be the only unbiased means of
identifying the primary formation epoch of massive galaxies 
(assuming such a unique epoch exists)
if the ISM of massive high-redshift galaxies is rapidly
enriched to at least solar metallicity. 
In the local universe, star-formation within 
the most massive metal-rich galaxies 
produces the highest rest-frame UV-extinction (Heckman 1998) and naturally, 
given a suitable dust covering factor, the galaxy luminosity is 
dominated by re-radiated 
FIR emission (L$_{\rm FIR}$/L$_{\rm UV} \sim 10 - 100$). 
Consequently one cannot necessarily expect 
optical studies of high-redshift galaxies ({\it e.g.} Steidel {\it et al.} 1996,
Madau {\it et al.} 1996) to provide an accurate
picture of the level of starforming activity in the young universe without
a more precise understanding of the impact of dust-obscuration on such 
rest-frame ultraviolet observations. 

New bolometer arrays recently commissioned on the largest 
sub-mm/mm telescopes (JCMT, CSO and IRAM) now have the 
sensitivity necessary to mount meaningful searches for a 
dust-enshrouded population of high-redshift prim\ae val galaxies, 
where we define prim\ae val to mean the progenitors 
of low-redshift massive ellipticals which have
gathered together the major fraction of their final baryonic mass and 
are caught early in the process of converting this material into stars.  

In this review we concentrate on describing 
the opportunities to make sub-mm
and millimetre continuum 
observations of high-redshift AGN with existing and future ground-based 
and space-borne instrumentation, summarise the new results from 
the recently commissioned bolometer array SCUBA 
(Gear \& Cunningham 1995), operating on the 15-m 
James Clerk Maxwell Telescope, and highlight the 
uncertainties that affect their interpretation.
We also briefly discuss the potential impact on models of galaxy 
formation/evolution of such pointed observations
of known high-redshift sources, and of the first deep sub-mm
surveys which are currently underway.
Molecular emission and absorption line investigations are described 
elsewhere in these proceedings.

Throughout this paper we assume $q_{0} = 0.5$ and 
$H_{0} = 50$\,kms$^{-1}$\,Mpc$^{-1}$ unless stated otherwise and correct all 
previously published data to the same cosmology.

\section{Observing dust emission at high redshift}

Ground-based astronomy is provided with a few extremely dry, 
high-altitude, stable observing
sites ({\it e.g.} Mauna Kea - Hawaii, Atacama Desert - Chile, 
South-Pole and Dome A - Antarctica) 
with sufficient atmospheric transmission that 
sensitive sub-mm and mm wavelength observations 
($350 - 1300\mu m$) can be attempted with broad-band continuum 
receivers. There is a great determination to exploit such high-quality sites.
New interferometric arrays (SMA, MMA, LMSA, LSA),  
large-single dish sub-mm 
(South Pole 10-m) and millimetre (LMT) telescopes, together with upgrades to the 
IRAM Plateau de Bure interferometer are all currently proposed,
and receiver development continues apace. 
Future 
space-borne sub-mm/FIR telescopes ({\it e.g.} SIRTF, FIRST) will have
the obvious advantage of complete coverage of the FIR--sub-mm regime, 
unhindered by 
atmospheric absorption, but the limited aperture size of such facilities
(often restricted to $< 4-$m), results in larger beam sizes and 
inevitably a greater vulnerability to galactic and extragalactic source 
confusion (Helou \& Beichman 1990, Gautier {\it et al.} 1992).
Within the next decade new ground-based, 
airborne and satellite telescopes will provide the powerful combination of 
large collecting area, sub-arcsec resolution,
significant instantaneous sky coverage (using large format 
($\sim 1000$ pixel) arrays)
and spectral coverage between $60\mu m - 3$mm, all of which are 
necessary if sub-mm astronomy is to continue to build 
on its recent impact on cosmological studies. 

\begin{figure}
\vspace{11.0cm}
\includegraphics{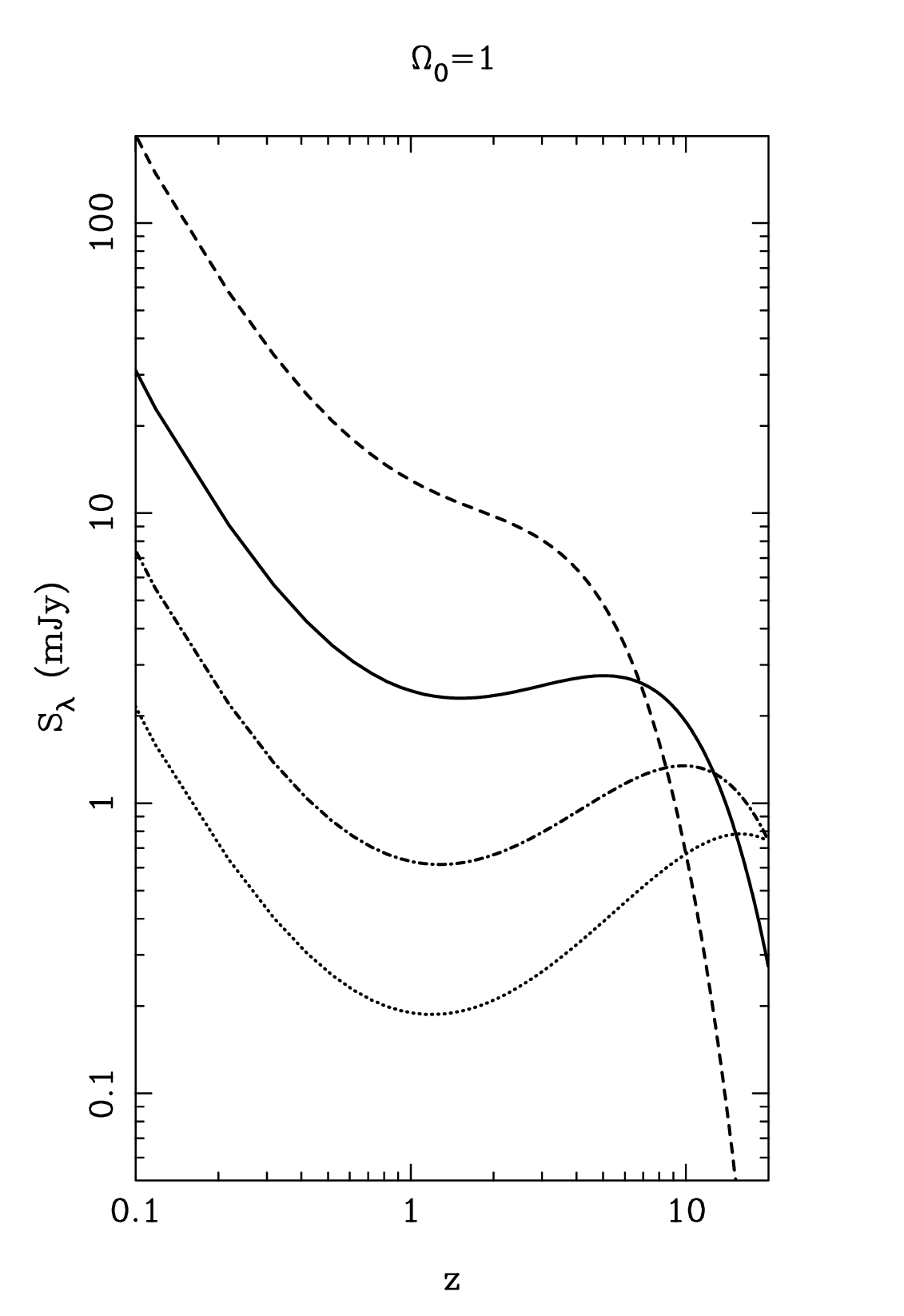}
\includegraphics{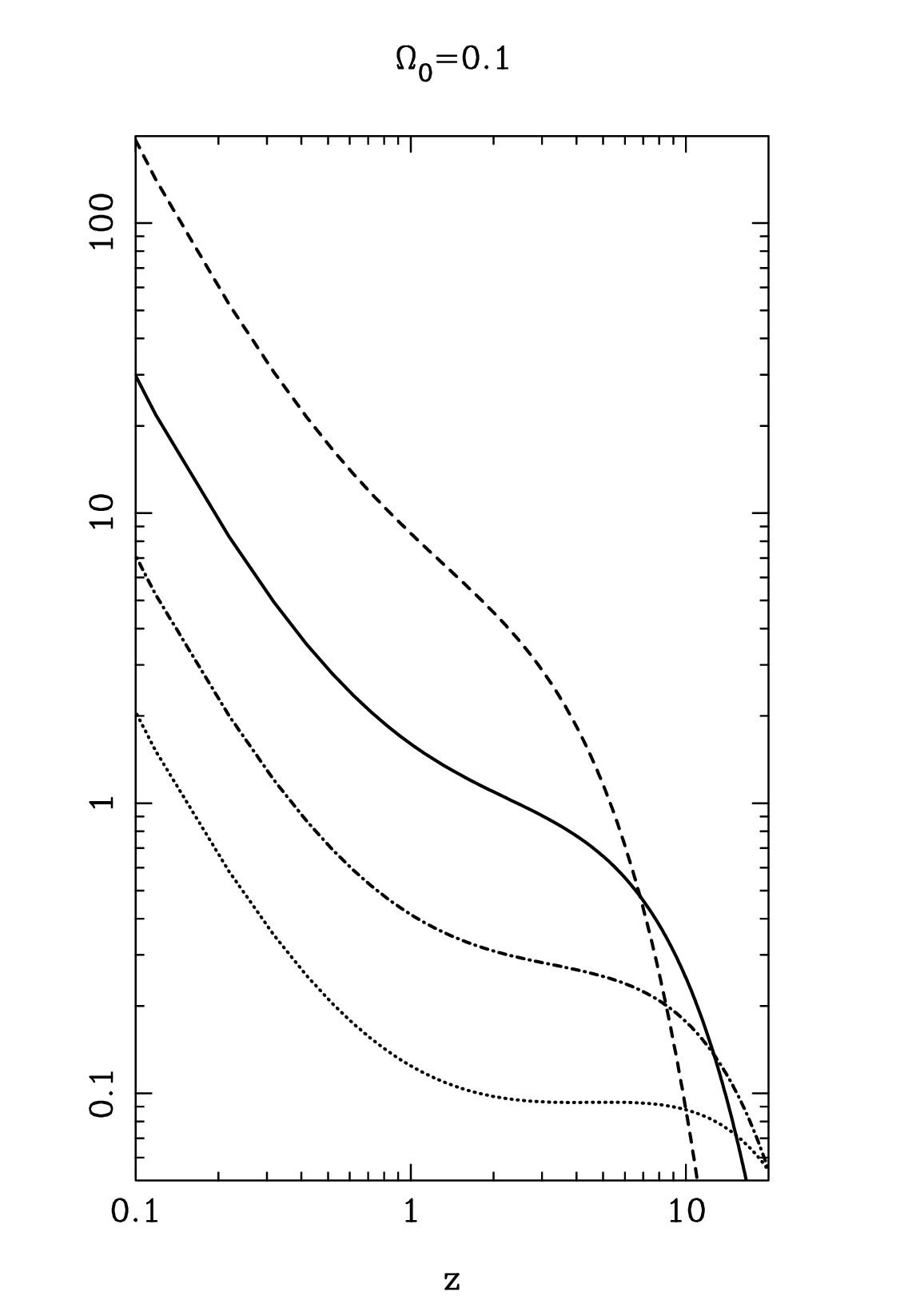}
\caption{\label{fig} 
The redshift dependence of flux density derived from the SED
of the starburst galaxy Arp220 
(z=0.018, ${\rm L_{FIR} \sim 3 \times 10^{12} L_{\odot}}$, 
${\rm SFR  \sim 300 M_{\odot}yr^{-1}}$) at 450$\mu m$ 
(dashed line), 850$\mu m$ (solid line), 1350$\mu m$ 
(dashed-dotted line) and 2000$\mu m$ (dotted line) as predicted for both
an Einstein-de Sitter
(LH panel) and low-$q_0$ (RH panel) universe.}
\end{figure}

The reason that sub-mm observations of starburst galaxies at 
cosmological 
distances are at all possible is illustrated in figure\,1. When attempting to observe most types of astronomical object out to extreme redshifts, observational cosmologists usually suffer the `triple whammy' of cosmological dimming,
a steeply declining luminosity function, and $+$ve k-corrections.
However, the steep spectral-index of the Rayleigh-Jeans 
emission from dust 
(${\rm F}_{\nu} \propto \nu^{2+\beta}, \beta \simeq 1.5$)
radiating at $30 - 70$K produces a negative 
k-correction at 850$\mu m$ 
of sufficient strength to completely compensate for cosmological
dimming beyond $z \simeq 1$, with the result that, certainly in an
Einstein-de Sitter universe, a dust-enshrouded starburst galaxy of a 
given luminosity
should be as easy to detect at $z \simeq 10$ as at 
$z \simeq 1$. The situation is inevitably 
less favourable for low values of $q_0$, but nevertheless the 
850\micron\ flux density is only expected to decrease by a factor of a few
between $z = 1$ and $z = 10$ before falling
dramatically at $z > 10$ as the redshifted peak of the SED 
($\lambda_{\rm rest} \sim 60-200\mu m$)
moves to longer wavelengths through the filter passband. 
This relative `ease' of access to the very high-redshift universe is
unique to sub-mm cosmology.

Figure\,1 also shows that the choice of the optimal observing wavelength 
for high-redshift studies has to be made with care.
Four competing factors influence the decision: (i) 
the observed-frame spectral shape,
which is a function of temperature, optical-depth and the 
frequency dependence of the grain emissivity; (ii) the presumed formation 
epoch of massive galaxies;
(iii) instrumental sensitivity, and (iv) the adopted cosmological model 
($\Omega_{0}$).

The sensitivities of bolometer instruments on all major sub-mm/mm facilities 
have been summarised by  Hughes \& Dunlop (1998) and Stark (1998) and 
indicate that observations at 850\micron\ with 
SCUBA, operating on the JCMT, 
currently provide the most sensitive combination of instrument, telescope,
observing wavelength and atmospheric conditions. 
For example, if the high-redshift universe contains starburst galaxies 
with FIR properties similar to Arp220 (z$= 0.018$, 
${\rm L_{FIR} = 3 \times 10^{12} L_{\sun}}$, ${\rm SFR 
\sim 300 M_{\sun}/yr}$), then at 850\micron , SCUBA can detect a similar 
galaxy at $z = 1 - 10$, with known position, in $\sim 3$\,hours, 
or can fully sample a 
random extragalactic field covering an area 
6 arcminutes$^{2}$ in $\sim 30$\,hours ({\it e.g.} 
the Hubble Deep Field, \S5)  
and identify active star-forming galaxies previously undiscovered through 
optical, IR or radio surveys (see figure\,2). 
Whilst in the absence of 
significant lensing amplification ($A > 10$)
it will continue to be impossible to detect {\em normal} galaxies
with moderate star formation rates ($\ll 10$\,M$_{\sun}$/yr) at $z > 0.5$, the 
new sub-mm/mm bolometer arrays are sufficiently sensitive that we can now
expect to observe a short-lived 
luminous formation phase of the most massive galaxies
which may exist at $z > 2$ provided that the SFRs exceed
$\sim 100$\,M$_{\sun}$/yr. Again we note that the required
integration times are significantly longer in an open universe.

The earliest sub-mm continuum observations of high-redshift
sources were attempts to detect known powerful radio galaxies and 
radio-quiet quasars at 800$\mu m$ (UKT14 $-$ JCMT) 
and 1300$\mu m$ (MPIfR 7-channel bolometer array $-$ IRAM 30-m), 
and although a few detections demonstrated the 
potential power of such observations ({\it e.g.} Barvainis {\it et al.} 1994, 
Dunlop {\it et al.} 1994, Chini {\it et al.} 1994, Isaak {\it et al.} 1994, 
Ivison 1995, Omont {\it et al.} 1996b, 
Hughes, Dunlop \& Rawlings 1997), many more continuum, 
but especially spectral-line 
observations, were unsuccessful 
for the possible reasons described below.

\begin{figure}
\vspace{11.0cm}
\includegraphics{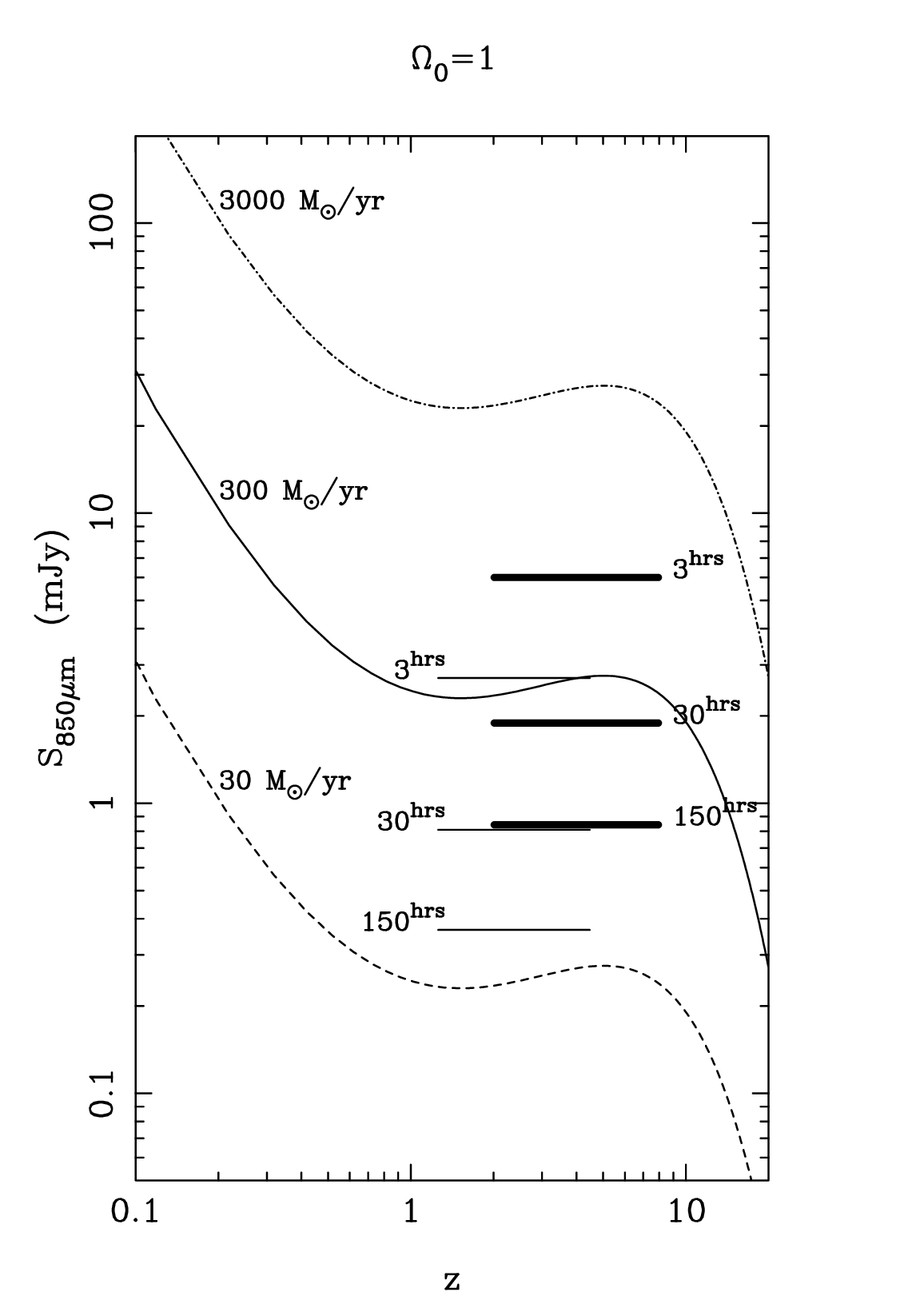}
\includegraphics{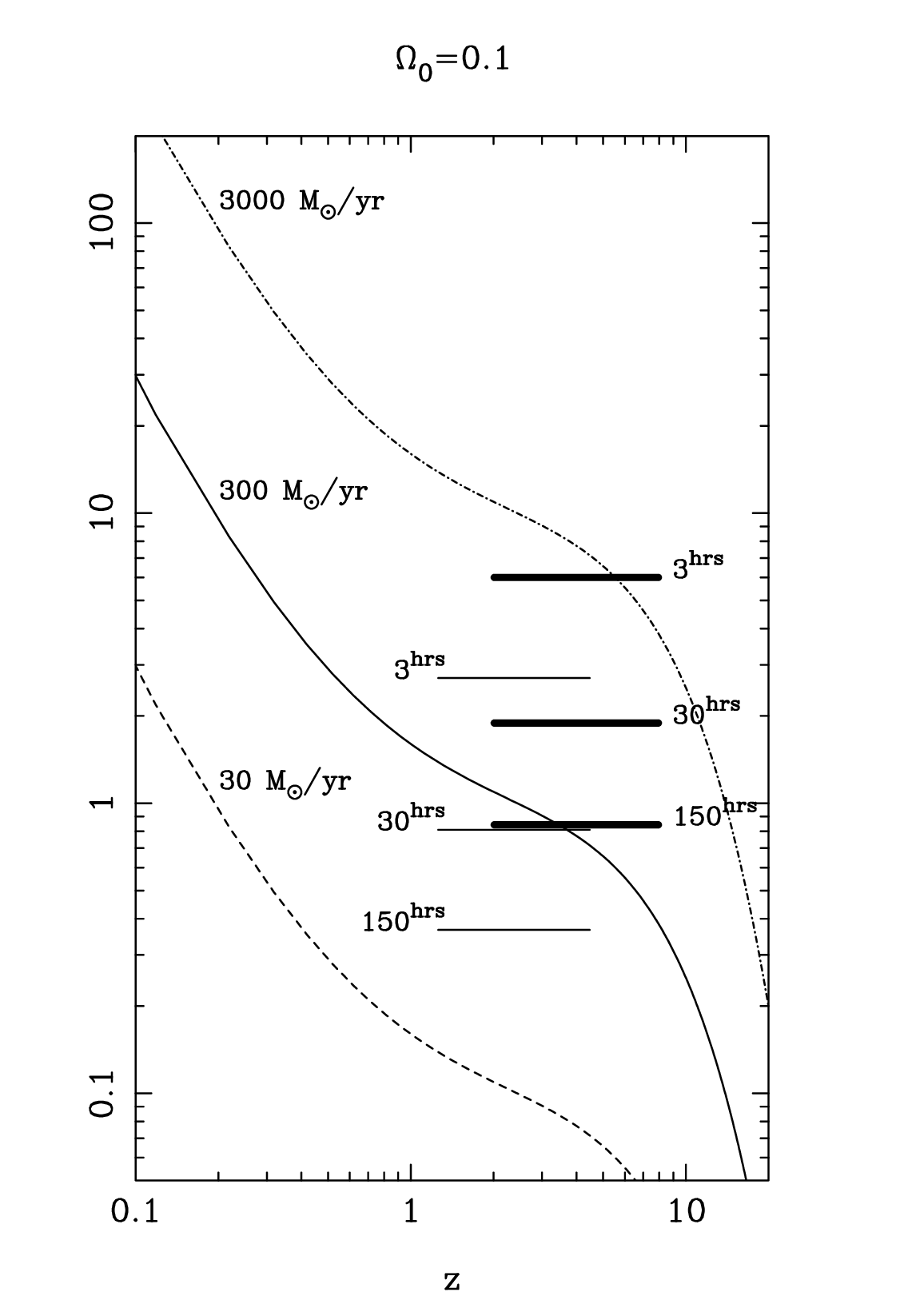}
\caption{\label{fig} 
The $S_{850\mu m} - z$ relationship normalised to the 850$\mu m$ 
flux-density of Arp220 (see figure\,1) which has a SFR $\sim 
300$M$_{\odot}$yr$^{-1}$ (solid curve).
Also shown are curves for a galaxy with a 
FIR luminosity and SFR $\times 10$ (dot-dashed 
curve) and $\times 0.1$ (dashed curve) that of Arp 220. The heavy and light 
horizontal lines show the 3-$\sigma$ flux densities currently 
achievable with SCUBA when used 
in MAPPING (a single fully-sampled pointing) 
and PHOTOMETRY modes in integration times of 3, 30 and 150\,hours respectively.
The predicted $S_{850\mu m} - z$ relationship is shown for two alternative cosmologies.}
\end{figure}

An assessment of the evolutionary status of the hosts of high-redshift AGN, 
based on these collective sub-mm/mm continuum data taken before 1997, 
was presented by Hughes, Dunlop \& Rawlings (1997) who
concluded that the FIR properties of high-redshift radio galaxies 
and radio-quiet quasars were more comparable with those of 
low-redshift interacting and star-forming ultraluminous FIR galaxies 
than with those expected of prim\ae val massive ellipticals, if such objects
formed the bulk of their present-day stellar populations in a single major star-burst at high-redshift (and thus might be expected to have 
SFRs $> 500 {\rm M_{\odot}/yr}$ and molecular 
gas masses ${\rm M_{H_{2}} > 5 \times 10^{11} M_{\odot}}$ - see \S 2). 

The apparently greater success rate of sub-mm
observations of high-redshift RQQs (Omont {\it et al.} 1996b), compared to 
that of radio galaxies, may be due to lensing which could
 have aided their initial identification in optical flux-limited samples, 
and which also amplifies their rest-frame FIR 
luminosity. It might also suggest that the hosts of high-redshift 
radio-quiet quasars are more gas rich than their radio-loud counterparts.
However, it may simply reflect the greater number of {\it 
extremely luminous} AGN available for study in optical quasar samples 
compared to
radio surveys because, interestingly, in both cases it is the objects of 
greatest luminosity (${\rm M_{B}} < -27$ in the case of QSOs, 
and ${\rm log(P_{408\,MHz}/WHz^{-1}sr^{-1}} 
> 27.5 $ in the case of radio galaxies - 
see next section) which have proved to be 
most easily detected at sub-mm wavelengths 
(Dunlop {\it et al.} 1994, Ivison {\it et al.} 
1998a, Omont {\it et al.} 1996b). If confirmed, such a correlation between sub-mm 
luminosity and AGN activity at high-redshift might arise for a number of reasons, including an underlying correlation with host galaxy mass (see \S4).

Some obvious possible explanations for the failure to detect large numbers of 
high-redshift active galaxies are that (i) elliptical galaxies do not 
form during the collapse of single massive halo with a luminous, 
but short-lived burst of starformation, and instead form through the hierarchical clustering 
and merging of lower-mass  gas clouds, which 
may already have formed their first generation of stars at an even 
higher redshift;  
(ii) that the formation  phase is more intense than expected, but with a 
proportionately shorter timescale, making it harder 
to ``catch one in the act''; 
(iii) that massive elliptical galaxies are more or less fully 
assembled, and the bulk of their stars have formed,
prior to $z \simeq 4$, and/or 
(iv) that we live in an open universe and our observations are therefore 
effectively less sensitive.

The imaging capabilities and greater sensitivity of SCUBA 
will alleviate most of these difficulties by allowing pointed 
observations of high-redshift AGN and wide-field surveys that 
cover a much larger region of parameter space (redshift, radio power, FIR 
luminosity, dust and gas mass). This will lead to a  
clearer understanding of the extent to which AGN hosts are reliable probes of the evolution of massive galaxies, and the 
relationship between the properties of the AGN and its host galaxy.
Furthermore the sub-arcsec resolution of future 
interferometers (where 1\,arcsec corresponds to $\sim 6$\,kpc at ${\rm z} = 4$) 
may enable follow-up observations to 
discriminate whether the source of the FIR emission originates 
from a single luminous galaxy, or multiple merging galaxies. 
In the remainder of this paper we describe the first observations of 
high-redshift radio galaxies with SCUBA, and the prospects for the first deep
`blank field' surveys which, at the time of writing, we
are currently undertaking.

\section{SCUBA observations of high-redshift radio galaxies}

\subsection{Study design}
An initial step towards an understanding of the FIR properties of the 
galaxies in the first few Gyr of the universe ({\it i.e.} at ${\rm z} > 2$)
has been to make pointed observations of known high-redshift 
active galaxies (radio galaxies and radio-quiet quasars). 
The criteria used to determine which sources are selected as 
prime candidates for sub-mm observations 
are mixed, and include strong ultra-steep 
spectrum radio emission ({\it e.g.} Dunlop {\it et al.} 1994, Hughes, Dunlop \& Rawlings 1997),
bright optical emission ({\it e.g.} 
Omont {\it et al.} 1996a,b),  
and high lensing amplification (Barvainis {\it et al.} 1994). 
An exciting development has been the recent sub-mm imaging of two 
intermediate-redshift lensing clusters (Smail, Ivison \& Blain  1997), which 
provide a small amplification ($A \simeq 1.5 - 3$) of the high-redshift 
universe, and which has in turn resulted in the discovery 
of an ultraluminous active galaxy at $z \sim 2.8$ (Ivison {\it et al.} 1998b).

\begin{figure}
\vspace{9.0cm}
\includegraphics{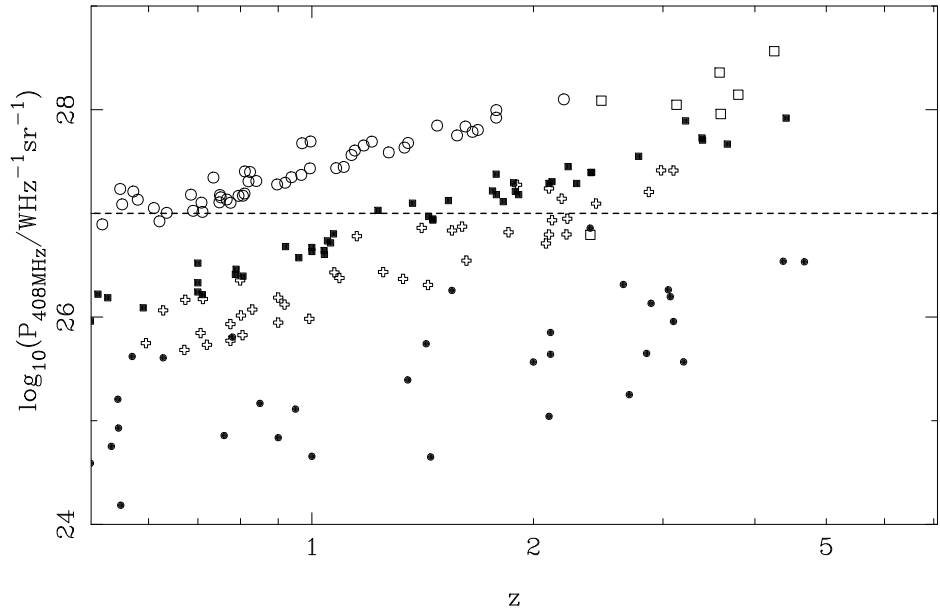}
\caption{
The radio-luminosity:redshift ($P - z$) plane of high-z galaxies selected from
progressively deeper radio surveys (3C - open circles; 6C - filled squares; 
7C - crosses; LBDS - filled circles) to be observed at sub-mm
wavelengths.
Using sub-mm observations of these samples we can 
quantify the contribution of an AGN to the rest-frame FIR   
luminosity, and trace the evolution of gas mass and star-formation rate
as function of redshift and radio luminosity.
The dashed line delineates the region of 
parameter space  log$(P_{\rm 408 MHz}/{\rm WHz^{-1}sr^{-1}}) > 27.0$ 
to which we are currently confining our SCUBA observations, in order to be able to quantify the cosmological evolution of dust and gas in radio galaxies of fixed radio luminosity.
The sub-mm data for the subset of radio galaxies observed to date
are presented in table\,1. 
} 
\end{figure}
 
We have concentrated on studies of steep-spectrum radio galaxies 
primarily because, as mentioned above, their low-redshift counterparts
are exclusively massive ellipticals and object selection on the basis 
of extended radio emission should be essentially free from lensing bias.
Thus, while radio galaxies may not be the easiest high-redshift 
objects to detect at sub-mm wavelengths, there is a genuine prospect
that their sub-mm emission (or lack of it) can provide some meaningful 
constraints on the evolution of massive ellipticals. Consequently
our collaboration is currently endeavouring to build 
on the work of Dunlop {\it et al.} (1994) 
and Hughes, Dunlop \& Rawlings (1997), and to attempt to address the 
question 
``{\em how and when did massive elliptical galaxies form?}'' 
by conducting a systematic 850$\mu m$ survey 
of radio galaxies spanning the redshift range  $z \simeq 0 - 5$ 
and a wide range of radio power ($24.0 < {\rm log(P_{408 MHz}/ 
W Hz^{-1}sr^{-1}}) < 28.5$). With  observations of  
a large sample of steep-spectrum, 
extended radio galaxies selected from progressively deeper
low-frequency radio surveys (3C, 6C, 7C, LBDS)  
we can avoid non-thermal contamination of the sub-mm emission, minimize 
any bias in the interpretation due to lensing, and break the correlation 
between redshift and radio luminosity
which arises in a survey with a single radio flux limit (Figure 3).

\subsection{Uncertainties in derived physical parameters}

In principle it is a straightforward exercise to derive 
the relevant physical parameters for high-redshift galaxies 
(such as rest-frame FIR luminosity, SFR, dust mass, molecular gas mass)   
from sub-mm continuum data, but in practice one has to proceed with 
caution. The first step is to use, if possible, 
a measure of the sub-mm spectral index to 
reject self-absorbed non-thermal synchrotron emission
in favour of thermal re-radiation from dust grains as the source of the FIR 
luminosity (Chini {\it et al.} 1989, Hughes {\it et al.} 1993). Second one has to establish that this thermal emission is unambiguously associated with the source, and not, for example, with galactic cirrus.
Third there are a number of properties related to
the grains and their emission that must be constrained (such as grain size, 
density, temperature distribution and frequency dependence of emissivity). 
Fourth, one must be able to exclude or quantify any gravitational lensing 
amplification. Fifth one has to be aware of the effect of an uncertain 
cosmology. Sixth, to convert dust mass into total gas mass requires assumptions
to made about, or constraints to be placed upon, the molecular gas:dust ratio
and also the metallicity of the ISM in massive galaxies at early epochs
(Eales \& Edmunds 1996).
A full discussion of all of these uncertainties, and how at least some of them can best be minimized is given by 
Hughes, Dunlop \& Rawlings (1997).

If one assumes or can establish that the sub-mm continuum 
($\lambda_{rest} > 200\mu m$)  is due to optically-thin emission 
from heated dust grains with no additional contribution from 
bremsstrahlung or synchrotron radiation,
a  measure of the dust mass $M_{d}$ can be determined directly from
the  relationship,

\begin{equation}
 M_{d} = \frac{1}{(1+z)} \frac{S_{obs} D_{L}^{2}}{k_{d}^{rest}
B(\nu^{rest},T)} 
\end{equation}

\noindent 
where $z$ is the redshift of the source,
$S_{obs}$ is the observed flux 
density, 
$k_{d}^{rest}$ is the rest-frequency mass absorption coefficient,
$B(\nu^{rest},T)$ is the rest-frequency value of the Planck function
from dust grains radiating at a temperature $T$, and 
the luminosity distance, $D_L$, is given by 
 
\begin{equation}
D_{L} = \frac{2c}{H_0 \Omega_0^2} \left\{\Omega_0 z + (\Omega_0 -
2)[(\Omega_0 z + 1)^{1/2} - 1] \right\}
\end{equation}

\subsection{Results}
We have recently commenced this major SCUBA programme by 
concentrating on observations of radio galaxies in the highest decade of radio 
luminosity for which radio galaxies can be found at all redshifts $z = 0.5 - 5$
({\it i.e.} objects in our sample with 
${\rm log(P_{408 MHz}/W Hz^{-1}sr^{-1}}) > 27.0$ - see figure 3). 
The subsample observed to date is listed in Table 1 which gives 
redshift, radio power, and the new measurements of
850$\mu m$ flux density for each galaxy. We also give our best estimates of 
the physical quantities which can be derived directly from the 850$\mu m$ measurement -- FIR luminosity, SFR and dust mass. 
The FIR properties of these high-redshift radio galaxies
are compared against the properties of local starforming galaxies in figure\,4.
These new SCUBA data reinforce our earlier conclusion that many of these 
high-redshift ellipticals 
display much more intense starforming activity and contain much larger
masses of dust and gas than their low-redshift 
counterparts (Chini {\it et al.} 1989, Knapp \& Patten
1991, Hughes {\it et al.} 1993).
However, figure 4 also demonstrates that while unusually active in the FIR 
compared to low-redshift radio galaxies and other ellipticals, very few 
high-redshift radio galaxies appear to be forming stars at a rate significantly higher than is found in low redshift ULIRGS. This appears to also be true 
for radio-quiet quasars, particularly if one allows for the possibility of
significant lensing of FIR emission from extreme objects such as 
BR1202$-$0725 
(Ohta {\it et al.} 1996, Omont {\it et al.} 1996a). Thus, with the possible exception of 4C41.17 and 8C1435+635,
none of the high-redshift radio galaxies we have observed to date appears 
to have the extreme FIR emission expected from a massive elliptical 
galaxy forming the bulk of its eventual stellar massive in a $\simeq 1-$Gyr 
star-burst.

\begin{table}
\caption{Physical parameters\tablenotemark{a}\ \
derived from 850$\mu m$ flux densities of high-redshift radio galaxies,
including radio power, FIR luminosity, starformation rate and dust mass.} 
\begin{center}\scriptsize
\begin{tabular}{llrrrrrr}

Source      &  \,\,\,\,z   & S$_{850\mu m}$ & $\Delta S_{850\mu m}$ 
& log P$_{\rm 408\,MHz}$  & log L$_{\rm FIR}$\tablenotemark{b}  
& SFR\tablenotemark{c}\phantom{00}  &  ${\rm log M_{dust}}$ \\
            &       & (mJy)\phantom{0} & (mJy)\phantom{0} & ${\rm (WHz^{-1}sr^{-1})}$
    &   ${\rm (L_{\odot})}$\phantom{0} &  ${\rm (M_{\odot}/yr)}$ & ${\rm (M_{\odot})}$\phantom{0}\\        

\tableline                     
            &       &       &      &        &             &         &         \\                                          
53W002      & 2.39 &   0.9\phantom{0}  &  1.1\phantom{0}  &  26.79\phantom{0000} &  $<$12.83   &  $<$674\phantom{0}  & $<$8.20\phantom{0}  \\
3C340       & 0.78  &   1.0\phantom{0}  &  1.2\phantom{0}  &  27.10\phantom{0000}  &  $<$12.91   &  $<$810\phantom{0}  & $<$8.28\phantom{0}  \\
3C277.2     & 0.76  &   1.8\phantom{0}  &  1.1\phantom{0} &  27.17\phantom{0000}  &  $<$12.87   &  $<$734\phantom{0}  & $<$8.24\phantom{0}  \\
6C0901+35   & 1.91 &$-$1.9\phantom{0}  &  1.2\phantom{0} &  27.18\phantom{0000}  &  $<$12.90   &  $<$803\phantom{0} & $<$8.28\phantom{0} \\
3C217       & 0.89  &   0.2\phantom{0} &  0.8\phantom{0} &  27.32\phantom{0000}  &  $<$12.75   &  $<$562\phantom{0} & $<$8.12\phantom{0} \\
6C1204+37   & 1.78  &   0.8\phantom{0} &  1.0\phantom{0} &  27.38\phantom{0000}  &  $<$12.84   &  $<$684\phantom{0} & $<$8.21\phantom{0} \\
6C0930+38   & 2.39  &   1.5\phantom{0} &  1.1\phantom{0} &  27.39\phantom{0000}  &  $<$12.82   &  $<$674\phantom{0} & $<$8.20\phantom{0} \\
3C265       & 0.81 &$-$1.1\phantom{0} &  1.0\phantom{0} &  -----\phantom{0000}  &  $<$12.83   &  $<$684\phantom{0} & $<$8.21\phantom{0} \\
6C0905+39   & 1.88  &   2.8\phantom{0} &  0.9\phantom{0} &  27.54\phantom{0000}  &     12.79   &     627\phantom{0} &    8.17\phantom{0} \\
6C0032+412  & 3.67  &   2.4\phantom{0} &  1.6\phantom{0} &  27.67\phantom{0000}  &  $<$12.90   &  $<$792\phantom{0} & $<$8.27\phantom{0} \\
B20902+34   & 3.39 &   1.9\tablenotemark{d}\phantom{0} &  1.5\phantom{0} &  27.72\phantom{0000}  &  
   &         &         \\
3C324       & 1.21 &   2.4\phantom{0} &  1.0\phantom{0} &  27.73\phantom{0000}  &  $<$12.86   &  $<$727\phantom{0} & $<$8.24\phantom{0} \\
6C0140+326  & 4.41  &   3.0\phantom{0} &  1.5\phantom{0} &  27.75\phantom{0000}  &  $<$12.83   &  $<$679\phantom{0} & $<$8.21\phantom{0} \\
3C241       & 1.62  &   3.1\phantom{0} &  1.5\phantom{0} &  27.88\phantom{0000}  &  $<$13.02   & $<$1052\phantom{0} & $<$8.39\phantom{0} \\
6C1232+39   & 3.22  &   3.9\phantom{0} &  1.0\phantom{0} &  27.89\phantom{0000}  &     12.84   &     689\phantom{0} &    8.22\phantom{0} \\
3C294       & 1.78  &   1.2\phantom{0} &  0.8\phantom{0} &  27.92\phantom{0000}  &  $<$12.74   &  $<$547\phantom{0} & $<$8.11\phantom{0} \\
3C257       & 2.47 &   3.9\phantom{0} &  1.1\phantom{0} &  28.08\phantom{0000}  &     12.89   &     784\phantom{0} &    8.27\phantom{0} \\
4C41.17     & 3.80 &  12.3\phantom{0} &  1.8\phantom{0} &  28.14\phantom{0000}  &     13.30   &    1996\phantom{0} &    8.68\phantom{0} \\
8C1435+635  & 4.25 &   8.3\phantom{0} &  0.7\phantom{0} &  28.56\phantom{0000}  &     13.10   &    1275\phantom{0} &    8.48\phantom{0} \\

\end{tabular}
\end{center}

\tablenotetext{a}{H$_{0}$ = 50\,kms$^{-1}$Mpc$^{-1}$, $\Omega = 1.0$}
\tablenotetext{b}{assumes isothermal 50K}
\tablenotetext{c}{${\rm SFR/(M_{\odot}yr^{-1}) \simeq 10^{-10} 
L_{FIR}/L_{\odot}}$}
\tablenotetext{d}{non-thermal flat-spectrum radio core dominates sub-mm emission}
\end{table}

\begin{figure}
\vspace{14.0cm} \includegraphics{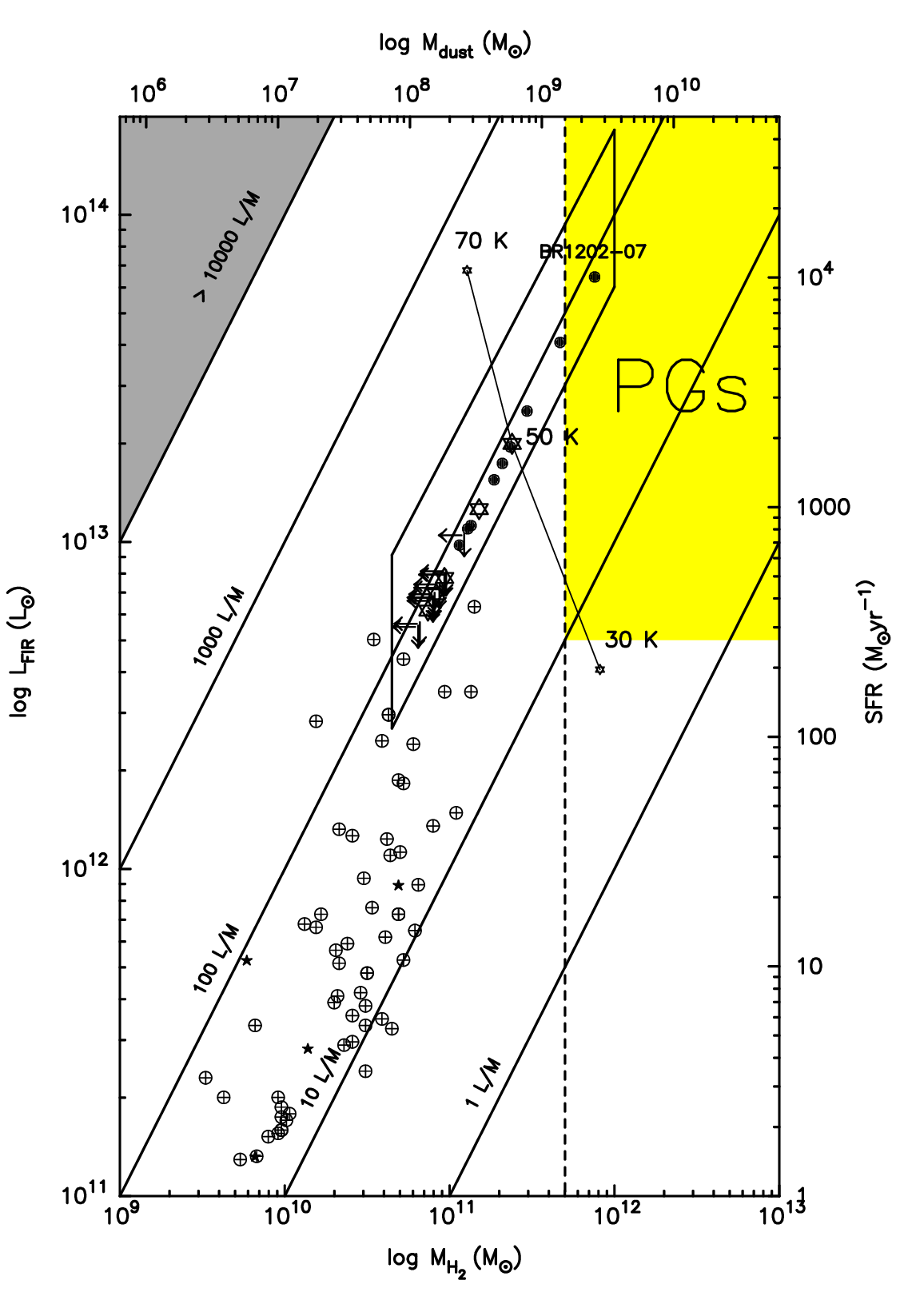}
\caption{ The physical properties of high-z radio galaxies (this paper - open
stars and limits) and radio-quiet quasars (Omont {\it et al.} 1996b - filled circles), 
which lie within the parallelogram, have been  derived from sub-mm or mm continuum
detections (assuming $T = 50$\,K,  M$_{H_{2}}$/M$_{d} = 500$) and are
compared to those of starburst galaxies (filled stars) and ULIRGs
(crossed circles) in the
local  universe. The diagonal lines indicate constant
L$_{FIR}$/M$_{H_{2}}$.  The vertical dashed-line shows the gas mass
boundary, to the right of which is a region of the parameter space
marked "PGs"  where one might expect to find the progenitors of the most
massive,  $> 5 \times 10^{11}$M$_{\odot}$, elliptical galaxies if they form 
by monolithic collapse.  This figure is adapted from Hughes, 
Dunlop \& Rawlings (1997).}
\end{figure}

Nevertheless, 4C41.17 and 8C1435+635 are undeniably interesting objects, 
in that they lie at extreme redshift ($z \simeq 4$) and
they are also the two most radio-luminous objects in our sample.
One of the most important questions to answer, therefore, is whether
their extreme sub-mm properties are primarily related to their extreme redshift
or their extreme radio power. 
Extension of our radio-galaxy survey to lower radio luminosities (at high-redshift) should help to answer this question. Ultimately, however, to properly quantify the level of star-formation at $z > 4$ requires unbiased 
blank-field sub-mm surveys, the first of 
which we now briefly discuss in the concluding section of this paper.

\section{Sub-millimetre Surveys}

Over the last few years giant strides have been made in the study 
of galaxy evolution/formation at optical/near-infrared wavelengths.
The completion of the first major galaxy surveys reaching $z \simeq 1$
({\it e.g.} Lilly {\it et al.} 1995), the development of sophisticated
spectral/dynamical/chemical models of galaxy evolution ({\it e.g.} 
Jimenez {\it et al.}
1998;
Baugh, Cole \& Frenk 1996, Kauffman \& Charlot 1998) and HST/Keck
observations of high-redshift galaxies ({\it e.g.} Pettini {\it et al.} 1997; Dunlop
{\it et al.} 1996, Ellis 1998) have combined to revolutionize the field.
 
Particular exciting has been the discovery of a substantial population of
high-redshift galaxies which has {\em not} been detected via nuclear
activity (Steidel {\it et al.} 1996). 
This has led to (arguably premature) attempts to delineate the
entire cosmological history of star formation out to $z \simeq 4$, with
initial analyses ({\it e.g.} Madau {\it et al.} 1996) suggesting that star-formation
activity peaked at relatively modest redeshifts $z \simeq 1 \rightarrow
1.5$ (a result hailed as a success for CDM-dominated models of
hierarchical structure formation).
However, such optically-based studies are always vulnerable to the effects
of dust which can lead both to 
under-estimation of star-formation rates in 
those UV-bright objects that are detected, and to 
potentially serious incompleteness at high redshift if a significant amount 
of star-formation takes place 
in dust-enshrouded environments.

\begin{figure}
\vspace{9.0cm}
\includegraphics{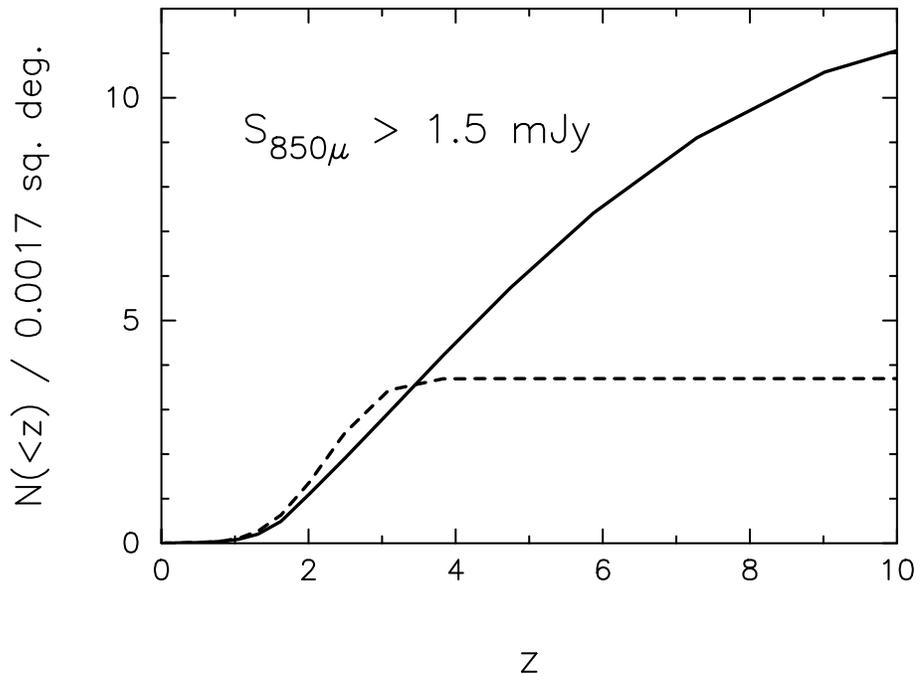}
\caption{\label{fig} 
Alternative example
predictions for the cumulative redshift distribution of sources detected
in a deep 850$\mu m$ image of the Hubble Deep Field reaching a complete
flux-density limit of 1.5 mJy. Two alternative predictions have been
produced by subjecting the local IRAS luminosity function to i) pure
luminosity evolution $(L(z) \propto (1+z)^3$ out to $z = 2$ with no
further evolution (either positive or negative) at higher redshifts
(solid line), and ii) evolution of the form displayed by powerful radio
galaxies, with a decline in space density beyond the peak at $z \simeq 2
-3$ (Dunlop \& Peacock 1990) .}
\end{figure}

The unbiased nature of radio-based selection at high-z,
coupled with the near-identical evolution of radio and star-formation
activity out to $z \simeq 1$, led Dunlop (1997) to suggest that the
evolution of radio activity should be regarded as the current best
predictor of the true evolution of star-formation activity. The
implication is that  starformation 
at $z > 2$ has been underestimated by a factor
$\simeq 5$ and interestingly factors of $3 \rightarrow 10$ have recently
been suggested by Pettini {\it et al.} (1997) and Heckman (1998) due to the
obscuring effect of dust on the UV light from starburst galaxies.
However even this may not be the complete story since the evolution
of radio-source activity may mirror a secondary phase of star-formation 
activity due to merging, with the primary epoch of star-formation 
lying at significantly higher redshift (as indicated by the 
large age of at least some high-z galaxies; Dunlop {\it et al.} 1996).
 
The key challenge in this field for the next few years is therefore to
properly quantify the level of star-formation activity in the high-z universe. 
If the large FIR luminosities of 4C41.17 and 8C1435+643
are in fact typical of massive galaxies at high redshift, then deep 
sub-mm surveys can be expected to contain large numbers of high-redshift galaxies.

At the time of writing we and our colleagues are performing the first 
deep SCUBA sub-mm survey of the sky, aimed at obtaining a 
deep 850$\mu m$ image of the Hubble Deep Field 
reaching a 3-$\sigma$ flux density limit of $\simeq 1 -1.5$ mJy. 
To whet the appetite we provide, in Figure 5, predicted cumulative redshift distributions for sources which should be detected in such a survey under two
alternative scenarios for the cosmological evolution of the dust-enshrouded
star-burst population. In the first case (dashed line) we have assumed that
the evolution of the IRAS galaxy luminosity function mirrors that of the
active galaxy radio luminosity function (as it certainly does out to $z \simeq 1$) at all redshifts, with the population peaking in luminosity at $z \simeq 2-
3$ and declining at higher redshifts (Dunlop \& Peacock 1990). Such a scenario
leads to a prediction of approximately 4 sources in the HDF with
$S_{850\mu m} > 1.5$mJy, all of which would be expected to lie at $z > 1$
but with none expected to lie at $z > 3$ due to the enforced high-redshift
`cutoff'. In the second case (solid line), 
a very similar level of luminosity evolution ($L(z) \propto (1+z)^3$) 
has been applied out to $z = 2$, with no further evolution in the luminosity
function (either positive or negative) at higher redshift.
Such a scenario would predict $\simeq 10$ sources, approximately half of which
could lie at $z > 4$. Clearly such an observation has the power to determine
whether or not substantial amounts of dust-enshrouded star-formation
occurred during the first Gyr of our Universe.

\acknowledgments

We thank our collaborators Steve Rawlings, Steve Eales and Elese Archibald 
for their continued contributions to this work.


%
%

\end{document}